\documentclass{epl}
\usepackage{amssymb}

\shorttitle{The equation of state of a hard-disc fluid}
\shortauthor{Brunner et al.}
\title{Measuring the equation of state of a hard-disc fluid}
\author{
M. Brunner \and C. Bechinger \and U. Herz \and H.H. von Gr\"unberg} 
\institute{
Universit\"at Konstanz, Fachbereich Physik, D-78457 Konstanz, Germany}
\begin{document}
\maketitle
\begin{abstract}
We use video microscopy to study a two-dimensional (2D) model fluid of
charged colloidal particles suspended in water and compute the
pressure from the measured particle configurations. Direct
experimental control over the particle density by means of optical
tweezers allows the precise measurement of pressure as a function of
density. We compare our data with theoretical predictions for the
equation of state, the pair-correlation function and the
compressibility of a hard-disc fluid and find good agreement, both for
the fluid and the solid phase. In particular the location of the
transition point agrees well with results from Monte Carlo
simulations.
\end{abstract}

Hard-disc (HD) fluids play a prominent role in liquid state theories.
This is due to the fact that, first, they often serve as reference
systems in perturbation theories of two-dimensional (2D) liquids (just
as hard-sphere fluids do for liquids in three dimensions), and that,
secondly, at high densities the behavior of every 2D fluid is
dominated by excluded volume effects, which in turn depends just on
the short-ranged hard-core part of the interparticle potential.
Mainly for theses two reasons, the HD equation of state (EOS) appears
also in many theories on monolayer adsorption on solid surfaces
\cite{steele76,steelebook}, an aspect illustrated for example in
Ref.~\cite{heimburg} where the HD EOS is used in statistical
mechanical theories modeling the binding of peripheral globular
proteins on lipid membranes. The important role of the HD system
explains the overwhelming number of theoretical studies on the EOS of
a HD fluid, starting as early as 1959 with the presentation of results
from scaled particle theory \cite{reiss59}.  Most of all approaches to
the EOS are based on particular resummations of the virial series and
the construction of sophisticated Pade approximants
\cite{oldEOS,henderson75,BAUS}, others use results from integral
equation theories \cite{intequ}, or from theories based on overlap
volume functions \cite{gonzalez91}.  Not only the EOS, but also the
structure of a hard-disc fluid has been explored in detail and is to
date well understood \cite{intequ,BAUS,gonzalez91,studart98}.  Hard
discs are popular also as model system to test advanced density
functional theories \cite{tarazona84,velasco97}, used, for instance, to
study discs in cavities \cite{kim01}, or induced freezing
and re-entrant melting \cite{bechinger01}.  Many computer
simulation studies of HD systems are available
\cite{steele76,oldSIM,newSIM}, of which the more recent ones have
focused mainly on the melting properties of HD solids \cite{newSIM}.

All these theoretical efforts contrast with the situation on the
experimental side. To our knowledge, the HD EOS has never been tested
with experimental data. The present Letter aims at closing this
gap. We report on experiments with a 2D model liquid of charged
colloidal particles suspended in water. Correlation functions computed
from real-space images, together with the virial equation, are used to
calculate the pressure $p$ of the 2D liquid as a function of the 2D
particle density $\rho$, which in our experiment can be conveniently
varied by means of optical tweezers. We have thus realized a 2D
colloidal model fluid for which the EOS, i.e. the $p(\rho)$-diagram,
can be determined directly. Comparing the experimental to the
theoretical EOS of HD's, we find the colloidal liquid to behave as a
2D fluid of HD's over a wide density range from the fluid to the solid
phase. 

We should remark that monolayers of atoms adsorbed on solid surfaces
behave in certain cases also as 2D fluids
\cite{steele76,steelebook}. In order to obtain experimental EOS's in
these atomic systems one has to transform measured isotherms to
$p(\rho)$ diagrams. This has been done by Glandt et al.
\cite{glandt79} who compared various theoretical EOS for a 2D
Lennard-Jones (LJ) fluid with the 2D pressure of argon and krypton,
adsorbed on graphitized carbon black \cite{thomy70}. Still, treating
adsorbed monolayers as 2D fluids remains an idealization which is
rarely justified because the 2D system is usually not isotropic due to
the natural corrugation of the substrate surface
\cite{steelebook,monson81,bechinger01}.

Contrary to atomic systems, where one is simply stuck with the
interaction dictated by the electronic structure of the atoms, the
interactions between colloidal particles can be externally controlled
and thus adapted to the problem one wants to study. To experimentally
realize a hard disc system with colloidal particles it is important to
assure that the inter-particle potential is extremely short ranged and
that no attractive parts in the potential exists. We have decided not
to work with sterically stabilized colloidal particles as this
stabilization usually leads to a structured pair potential in the
distance region were the polymer brushes start to overlap
\cite{marcus96}. Rather we chose to use charge-stabilized colloids at
moderate salt concentrations. A screening length $\kappa^{-1}$ between
$100\: nm$ and $200 \: nm$ yielded optimal results. Higher salt
concentrations resulted in an increased unfavorable particle substrate
interaction and a pronounced attractive van-der Waals contribution in
the interparticle potential. For the same reason it proved unpractical
to use very thin sample cells, as this is known to also induce an
attractive part in the pair potential \cite{grier01}. The colloidal
system employed consisted of charged sulphate-terminated polystyrene
spheres of $\sigma_0 = 3 \:\mu m$ diameter (IDC Cooperation) and of
charged sulphate-terminated silica particles of 2.4 $\mu m$ diameter
(polydispersity $< 4 \%$). The suspension was injected into a sample
cell made of fused silica plates with $200\: \mu m$ spacing. We
performed measurements at high and at low salt concentration, i.e. at
$\kappa \sigma_0 = 20.0$ and $\kappa\sigma_0= 6.9$, using the $2.4\:
\mu m$ spheres in the high-salt and the $3\:\mu m$ spheres in the
low-salt measurement.  In order to confine the particles to two
dimensions the widened beam of a frequency-double Nd:YV0$_4$ laser was
directed from above into the sample cell, exerting a vertical light
pressure on the particles and pushing them towards the bottom plate of
the sample cell. The light pressure and gravity on the one side and
the electrostatic particle-wall repulsion on the other side create a
sharp potential minimum in the vertical direction in which the
particles are trapped and thus effectively confined to two
dimensions. We have varied the intensity of the laser beam over a wide
range and observed no influence on the particle structure. Therefore
light induced effects on the pair interaction can be ruled out. In
contrast to the above-mentioned 2D LJ fluids, realized by adsorbed
atoms on solid surfaces, we here do not need to make special
assumptions concerning the properties of the substrate surface.  It
can be safely ruled out in our experiment that the particles' mobility
in 2D is in any way hindered by a possible periodicity of the
substrate surface. Our system is almost perfectly two-dimensional with
vertical out-of-plane fluctuations of less than $250\: nm$. An
important component of our set-up which is indispensable for measuring
$p(\rho)$ diagrams, is our method to vary the particle density
$\rho$. This was achieved by a scanned optical laser tweezers briefly
described in the following. The beam of a laser was reflected from a
computer-controlled system of mirrors and focused into the 2D system
plane. By means of these mirrors the tweezers was repeatedly (200Hz)
scanning a line having the form of a rectangular box, which results in
an optical line trap for the particles along the contour of the
box. Trapped particles arrange like a pearl-necklace along this
contour and, due to the repulsive inter-particle potential, prevent
other particles from escaping the box, thus defining a system (all
particles in the box) with a certain number of particles (1000 to
3000).  The box size can be changed via the control unit of the mirror
system which allows precise adjustment of the particle density in the
system. For each system at density $\rho$, digital video microscopy
measurements were made using a high-aperture long-distance microscope
objective (Zeiss Achroplan 63x) with dark field illumination. The
particle positions were determined with an accuracy of better than
$50\: nm$. From statistical averages of the particle positions, we
computed pair correlation functions $g(r)$. Further details about the
experimental set-up and the data analysis can be found in our
previous papers \cite{klein,brunner}.

Having the pair-correlation function at density $\rho$, the pressure
can be computed from the virial equation which for a
isotropic 2D system reads 
\begin{equation}
\label{eq:1}
\beta p \sigma^2 = \rho \sigma^2 - \frac{\pi \rho^2 \sigma^2}{2} \int_0^\infty dr r^2
\beta u^{\prime}(r) g(r) \:.
\end{equation}
where $\beta = 1/kT$. $u^{\prime}(r)$ is the derivative of the
interparticle potential $u(r)$ with respect to the distance $r$; this
potential is not known {\em a priori}, but may be derived from the
structural data, e.g. from inverting $g(r)$. This was done by means
of the Ornstein Zernike equation and appropriate closure relations
(Percus-Yevick and HNC), a method described and tested in
\cite{klein}, see also \cite{hidalgo}.  Performing then this integral
for all measured $g(r)$'s, we obtain $p(\rho)$, the desired EOS.
For the $\kappa \sigma_0 =20.0$ measurement, we obtained from the
inversion almost perfect Yukawa potentials ($u(r) \sim e^{-\kappa
r}/r$) which were -- within the error bars specified below --
identical, regardless at what density we analyzed the pair-correlation
function, while for the low salt measurement we obtained, for all
$g(r)$'s, Yukawa potentials with almost identical prefactors and
screening constants $\kappa$, but with a density-dependent truncation
at large distances and high densities. This density-dependent
truncation has been interpreted in terms of many-body interactions
\cite{brunner,klein}. The fact that in both measurements the $\kappa$
resulting from the inversion procedure do not show any dependence on
$\rho$, clearly indicates that the salt-ions dominate the screening
behavior and that the contribution of counter-ions to the screening
can be neglected. Due to the double-layers around the particles, the
effective hard-core diameter $\sigma$ is larger than the actual
particle diameter $\sigma_0$. Similar to \cite{barker67}, we
determined $\sigma$ by first evaluating the second virial coefficient
\begin{equation}
B_2 = - \pi \int_0^\infty r (e^{-\beta u(r)} -1) dr 
\end{equation}
using the interparticle potential $u(r)$ obtained from the measured
$g(r)$'s, and by then identifying $B_2$ with the second virial
coefficient of a pure HD system, $B_2^{(HD)} = \pi \sigma^2/2$. We
obtain $\sigma=1.084 \sigma_0$ for the $\kappa \sigma_0 =20.0$ measurement,
and $\sigma = 2.16 \sigma_0$ for the $\kappa \sigma_0 = 6.9$ measurement.
\begin{figure}
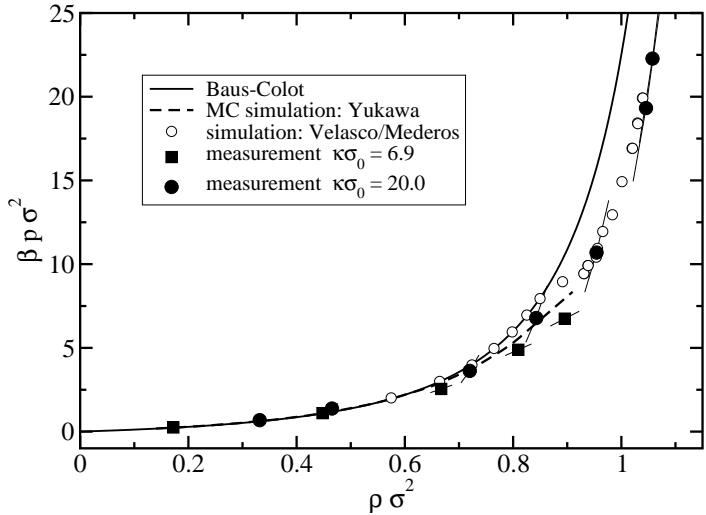

\onefigure[width=0.65\textwidth]{fig1.eps}
\caption{
Equation of state of a hard-disc system in the fluid and the solid
density range. Theoretical predictions for the fluid branch
(Baus/Colot \cite{BAUS}) and the solid
branch (Velasco/Mederos \cite{velasco97}) are compared with
experimental data, measured in high-salt (filled circles) and low-salt
(filled squares) colloidal suspensions.  Error-bars (solid short lines
attached to the filled symbols) are inclined for reasons explained in
the text. Monte Carlo data for a Yukawa fluid (dashed line) are
provided to interpret the $\kappa \sigma_0 = 6.9$ measurement.
}
\label{fig1}
\end{figure}

Fig.~(\ref{fig1}) shows the pressure as a function of $\rho \sigma^2$,
for both measurements. The precision of our procedure is limited
mainly by errors made in the determination of $u(r)$; they lead to
small variations in the computed pressure, but also in the effective
hard-core radii. Therefore, the error bars appear tilted in
Fig.~(\ref{fig1}). The measured EOS is compared with Monte-Carlo (MC)
data from \cite{velasco97}, and with an expression
proposed by Baus and Colot \cite{BAUS} for the EOS of a fluid HD
system,
\begin{equation}
\label{eq:3}
\beta p /\rho = \frac{1 + \sum_{n=1}^6 c_n \eta^n}{(1-\eta)^2}
\end{equation}
with $\eta = \pi \rho \sigma^2/4$ and $c_1=0, c_2=0.128, c_3= 0.0018,
c_4 = -0.0507, c_5 = -0.0533, c_6 = -0.0410$. The agreement between
our high-salt data and the HD EOS is almost perfect, both on the fluid
and the solid side of the phase transition, but also with regard to
the location of the transition point itself. A much simpler EOS
\cite{henderson75} ($c_2=0.125$ and $c_n=0$ for $n=1$, $n>2$ in
eq.~(\ref{eq:3})) agrees equally well and also the density-functional
theory by Velasco and Mederos \cite{velasco97} (data not shown
here). Surprisingly, even for low salt concentration ($\kappa \sigma_0
= 6.9$, $\sigma/\sigma_0 = 2.16$ !), the colloidal system can still be
successfully mapped to a HD fluid, at least for low densities.  For
$\rho \sigma^2 > 0.7$, there are marked deviations from the HD EOS. We
performed MC simulations to compute pair-correlation functions of a
quasi-2D fluid system, using the Yukawa part of $u(r)$ of the $\kappa
\sigma_0 =6.9$ measurement. The pressure curve, computed with
eq.~(\ref{eq:1}) from the MC-generated pair-correlation functions, is
plotted as dashed line in Fig.~(\ref{fig1}). It is evident that this
line differs, as expected, from the HD EOS, but also from the
experimental values of the $\kappa \sigma_0 =6.9$ measurement, a
difference which hence must be due to deviations of the experimental
pair potential from the assumed Yukawa form. As pointed out in
\cite{klein,brunner} the experimental pair potential shows a
density-dependent truncation in $u(r)$, which has been ignored in the
MC simulation. 

\begin{figure}
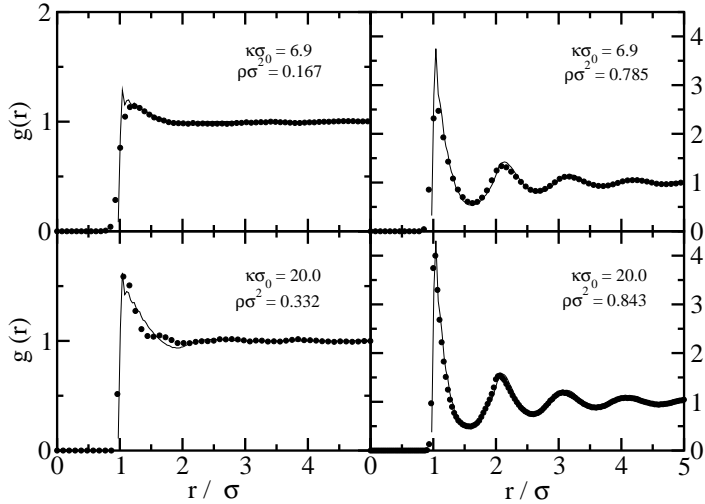

\onefigure[width=0.65\textwidth]{fig2.eps}
\caption{Comparison between theoretical \cite{BAUS} (solid line) and
experimental (filled circles) pair-correlation functions for a
hard-disc fluid at different densities, experimentally realized with
low-salt ($\kappa \sigma_0 =6.9$) and high-salt ($\kappa \sigma_0
= 20.0$) colloidal suspensions.}
\label{fig2}
\end{figure}

Baus and Colot \cite{BAUS} have suggested a semiempirical expression
for the direct correlation function of a HD fluid which can be related
to $g(r)$ via the Ornstein-Zernicke equation. Fig.~(\ref{fig2}) shows
the resulting pair-correlation functions (solid lines) for various
densities and compares them with our measured $g(r)$. We emphasize
that after matching the second virial coefficient no free parameter
was used. While in the case of high salt concentration excellent
agreement is found, differences are observed for the $\kappa \sigma_0
=6.9$ measurement, mainly for small particle separations where the
first peak is less pronounced compared to that of the HD fluid, due to
the softer pair potential. For larger distances, packing effects
dominate, and the agreement is again good.

\begin{figure}
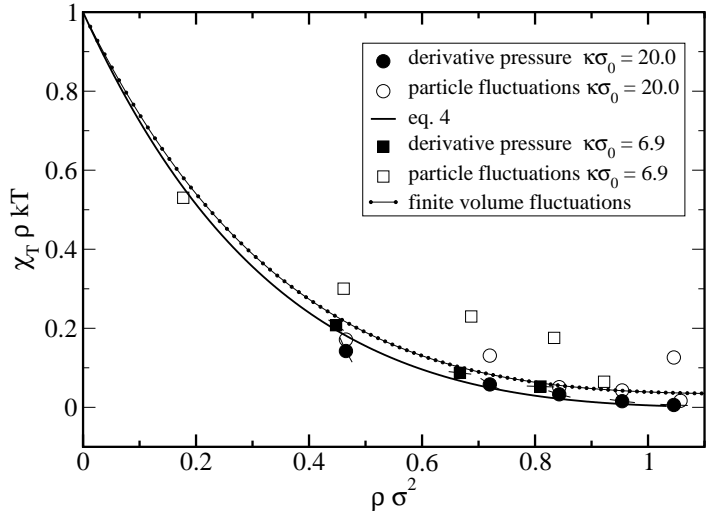

\onefigure[width=0.65\textwidth]{fig3.eps}
\caption{Compressibility of a hard-disc fluid, as obtained from
density derivatives of the $p(\rho)$ curves of Fig.~(\ref{fig1}) and 
the particle number fluctuations $\langle \Delta N^2\rangle/\langle N
\rangle$. Symbols for the experimental values, lines for theoretical
predictions for an infinite system, eq.~(\ref{eq:4}), and a finite
system \cite{roman97}.}
\label{fig3}
\end{figure}

From the recorded particle positions we can also compute the particle
number fluctuation $\langle \Delta N^2\rangle/\langle N \rangle$ with
$\Delta N = N- \langle N \rangle$, $\langle N \rangle$ being the mean
number of particles. This quantity which is related to the isothermal
compressibility, $\chi_T k T \rho = \langle \Delta N^2\rangle/\langle
N \rangle$, is plotted as open symbols in
Fig.~(\ref{fig3}). Alternatively, we can calculate $\chi_T$ from the
density derivative of the pressure, $\chi_T k T \rho = (\beta
\partial p/\partial \rho)^{-1}$, which we did in Fig.~(\ref{fig3})
(filled symbols) computing the differences between neighboring
experimental values in Fig.~(\ref{fig1}). The compressibilities,
calculated in both ways, are compared with the derivative of
Henderson's EOS \cite{henderson75}
\begin{equation}
\label{eq:4}
\rho kT \chi_T
= \Big[\frac{1+\eta^2/8}{(1-\eta)^2} + 
\eta \frac{\eta+8}{4(1-\eta)^3}\Big]^{-1} \:,
\end{equation}
plotted in Fig.~(\ref{fig3}) as solid line. While the density
derivatives of the experimental pressure values agree nicely with the
theoretical prediction of eq.~(\ref{eq:4}), $\langle \Delta
N^2\rangle/\langle N \rangle$ shows deviations, especially for the
low-salt measurement. One reason for this discrepancy might be a
finite size effect.  To estimate this effect, we used the
pair-correlation functions, suggested by Baus and Colot for the
infinite system, to calculate the particle number fluctuation in a
finite subvolume ($V = \pi r_0^2$, $r_0 = 10 \sigma_0$) of a hard-disc
fluid composed of a fixed number of particles (1000 particles), in a
way described in detail by Roman et al. \cite{roman97}. The
result is plotted in Fig.~(\ref{fig3}) and the small correction
illustrates that, at least for the high salt measurement, the observed
differences can be explained with a finite size effect. For the low
salt measurement, the remaining differences are probably due to insufficient
sampling. 

To obtain more information on the character of the different phases in
Fig.~(\ref{fig1}), we also examined the orientational correlation
function $g_6(r)$ \cite{nelson} for the high-salt measurement. Up to a
density of $\rho \sigma^{2} = 0.85$ the system exhibits a pure liquid
phase as confirmed by the isotropic pair-correlation function $g(r)$
and the exponentially decaying $g_6(r)$.  On the other hand, at
densities above $\rho \sigma^2 = 0.95$ the system is in stable
crystalline state, with a slow algebraically decaying pair correlation
function and a constant orientational correlation function. However,
in between, i.e. for $0.85 < \rho \sigma^2 < 0.95$, there is a
transition region, in which $g_6(r)$ decays slower than exponentially
and a local hexagonal order can be observed. At present it cannot be
decided whether this is the hexatic phase or a very polycrystalline
state. In contrast to the measurement of Marcus and Rice
\cite{marcus96} no pronounced liquid-hexatic and hexatic-solid
coexistence region has been observed, which is most likely due to the
fact that their pair-potentials had weak attractive parts, while ours
are purely repulsive.

We close with a few comments and summarizing remarks.  (i) Similar
hard-disc-like fluids have been experimentally realized also by Murray
et al.  \cite{murray87} and by Marcus et al. \cite{marcus96} using
colloidal suspensions in confined geometries. While these authors
concentrated on the melting properties of 2D systems, it has been the
focus of the present work to study the EOS of the 2D system which
naturally requires a convenient experimental tool to vary the density,
realized here with optical line tweezers. (ii) To experimentally
realize a HD system, it is decisive to choose the right salt
concentration, as salt tunes the inter-particle potential but also the
interaction between particle and wall. Too much salt leads to an
attraction in the interparticle potential and makes the wall-particle
interaction more short-ranged (so that the 2D system plane is shifted
too close to the wall), while too little salt results in a very soft
repulsive potential. In both cases, the colloidal system can no longer
be interpreted as a hard disc system. The extent of the deviation
between the experimental and the HD system caused by a very soft pair
potential ($ \kappa \sigma_0 = 6.9 $) has been shown above.  (iii) 2D
colloidal suspensions can indeed be considered as a hard-disc
system. This applies both to the structure as well as to the
thermodynamics. The effective HD radius of the colloidal particles is
correctly defined by simply demanding the second virial coefficient to
be equal in both descriptions. (iv) On the atomic level, the EOS of 2D
fluids can be derived from adsorption isotherms \cite{steelebook}. As
opposed to these experiments, we here have the full structural
information, controllable substrate-particle interactions and tunable
inter-particle potentials. In adsorbed monolayer films \cite{thomy70}
the phase behavior can be deduced only via averaged quantities and
there is a certain ambiguity about the emergence of multi-layer
adsorption at higher densities. This highlights once again the power
of colloids as a model system in statistical mechanics.

\end{document}